%% file: azimuth.tex
\documentclass[aps,prl,showpacs,twocolumn,groupedaddress]{revtex4}  
\usepackage{graphicx}  
\usepackage{dcolumn}   
\usepackage{bm}        
\usepackage{amssymb}   

\newcommand{\Dphi}{\Delta \phi\,{}_{\text{dijet}}}
\newcommand{\ptmax}{p_T^{{\text{max}}}}

\begin{document}

\title{\boldmath  
Measurement of Dijet Azimuthal Decorrelations at Central Rapidities \\
in $p\bar{p}$ Collisions at $\sqrt{s}=1.96\,\text{TeV}$ }
\input list_of_authors_r2.tex           
\date{\today}
\begin{abstract}
Correlations in the azimuthal angle between the two largest 
transverse momentum jets have been measured using the D\O\ detector 
in $p\bar{p}$ collisions at a center-of-mass energy 
$\sqrt{s}=1.96\,\text{TeV}$.
The analysis is based on an inclusive dijet event sample in the
central rapidity region corresponding to an integrated luminosity of
$150\,\text{pb}^{-1}$.  Azimuthal correlations are stronger at larger
transverse momenta.  These are well-described in perturbative QCD at
next-to-leading order in the strong coupling constant, except at large
azimuthal differences where soft effects are significant.
\end{abstract}

\maketitle

%
Multi-parton radiation is one of the more complex aspects of
perturbative Quantum Chromodynamics (pQCD) and its theory and
phenomenology are being actively studied for the physics programs at
the Fermilab Tevatron Collider and the CERN LHC~\cite{multiparton}.
The proper description of radiative
processes is crucial for a wide range of precision measurements as
well as for searches for new physical phenomena 
where the influence of QCD radiation is unavoidable.  
A clean and simple way to study radiative processes is to 
examine their impact on angular distributions. 
Dijet production in hadron-hadron collisions, in the absence of 
radiative effects, results in two jets with equal transverse momenta 
with respect to the beam axis ($p_T$) and correlated
azimuthal angles $\Dphi=|\phi_{\text{jet\,1}} - \phi_{\text{jet\,2}}| = \pi$.
Additional soft radiation causes small azimuthal decorrelations,
whereas $\Dphi$ significantly lower than $\pi$ is evidence of additional hard 
radiation with high $p_T$.
Exclusive three-jet production populates $2\pi/3 < \Dphi < \pi$
while smaller values of $\Dphi$ require additional radiation such as a 
fourth jet in an event.
Distributions in $\Dphi$ provide an ideal testing ground for 
higher-order pQCD predictions without requiring the reconstruction
of additional jets
and offer a way to examine the 
transition between soft and hard QCD processes based on a 
single observable.  

A new measurement of azimuthal decorrelations between jets produced at
high $p_T$ in $p\bar{p}$ collisions is presented in this Letter.  Jets
are defined using an iterative seed-based cone algorithm (including
mid-points) with radius ${\cal R}_{\text{cone}}=0.7$~\cite{run2cone}.  The same
jet algorithm is used for partons in the pQCD
calculations, final-state particles in the Monte Carlo event
generators, and reconstructed energy depositions in the experiment.  
$\Dphi$ is reconstructed from the two jets with highest $p_T$ in an event.
The observable is defined as the differential dijet cross section
in $\Dphi$, normalized by the dijet cross section integrated over 
$\Dphi$ in the same phase space
$(1/\sigma_{\text{dijet}})\, (d\sigma_{\text{dijet}}/ d\Dphi)$.
(Theoretical and experimental uncertainties are reduced in this
construction.)
Calculations of three-jet observables at next-to-leading order (NLO)
in the strong coupling constant $\alpha_s$, have recently become
available~\cite{nlo,nlojet}.  
Quantitative comparisons with data yield
information on the validity of the pQCD description and increase
sensitivity for gauging potential departures that could
signal the presence of new physical phenomena.

%
Data were obtained with the D\O\ detector~\cite{run2det} in Run~II of
the Fermilab Tevatron Collider using $p\bar{p}$ collisions at
$\sqrt{s}=1.96\,\text{TeV}$.  The primary tool for jet detection was a
compensating, finely segmented, liquid-argon and uranium calorimeter
that provided nearly full solid-angle coverage.  Calorimeter cells
were grouped into projective towers focused on the nominal interaction point
for trigger and reconstruction purposes.  Events were acquired using
multiple-stage inclusive-jet triggers.  
Four analysis regions were defined based on the jet with largest
$p_T$ in an event ($\ptmax$) with the requirement that the trigger
efficiency be at least 99\%.
The accumulated integrated luminosities for events with 
$\ptmax > 75$, $100$, $130$, and $180$~GeV were 
$1.1$, $21$, $90$, and $150\,\text{pb}^{-1}$ ($\pm6.5\%$), respectively.  
The second leading $p_T$ jet in each event was required to have 
$p_T>40$~GeV and both jets were required to have central rapidities with 
$|y_{\text{jet}}| < 0.5$ where 
$y_{\text{jet}}=\frac{1}{2}\ln{\bf (}(E+p_z)/(E-p_z){\bf )}$  and 
$E$ and $p_z$ are the energy and the longitudinal momentum of the jet.

\begin{figure}
\includegraphics[scale=0.96]{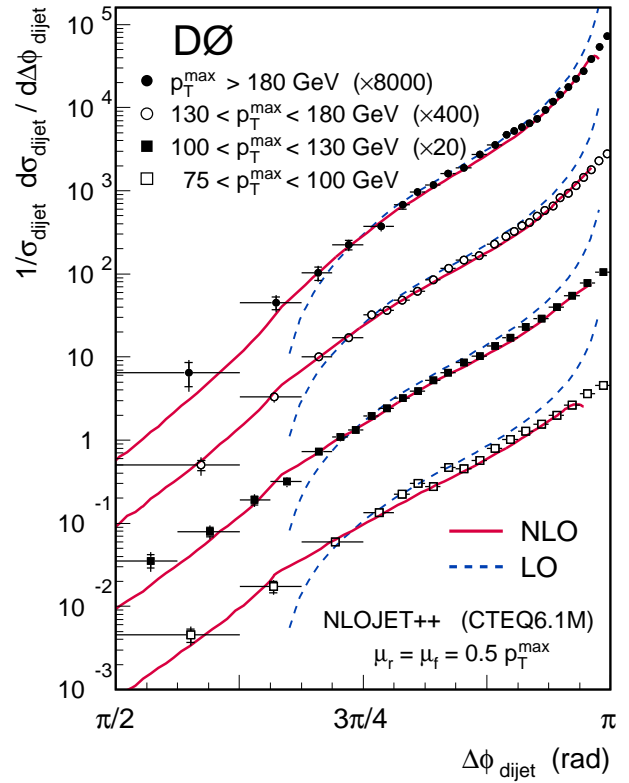}
\caption{\label{fig:data1} 
The $\Dphi$ distributions in four regions of $\ptmax$.  
Data and predictions with $\ptmax > 100\,\text{GeV}$ are scaled 
by successive factors of 20 for purposes of presentation.
The solid (dashed) lines show the NLO (LO) pQCD predictions.}
\end{figure}

The position of the $p\bar{p}$ interaction was reconstructed using a
tracking system consisting of silicon
microstrip detectors and scintillating fibers located within a $2\,\text{T}$
solenoidal magnet.  The vertex $z$-position was required to be within
$50$\,cm of the detector center which preserved the projective nature
of the calorimeter towers.  
The systematic uncertainty associated with the vertex selection efficiency 
is less than $3\%$ for $\Dphi>2\pi/3$ and
$\approx8\%$ for $\Dphi\approx\pi/2$.  
The missing transverse energy was calculated from the vector sum of 
the individual transverse energies in calorimeter cells.  
Background from cosmic rays and incorrectly vertexed events was eliminated by 
requiring this missing transverse energy to be below $0.7\,\ptmax$.  
Background introduced by electrons, photons, and detector noise that 
mimicked jets was eliminated based on characteristics of 
shower development expected for genuine jets.
The overall selection efficiency is typically $\approx 83\%$ for 
$\Dphi < 5\pi/6$ and drops to $\approx 78\%$ as $\Dphi\rightarrow \pi$.

The $p_T$ of each jet was corrected for calorimeter showering effects,
overlaps due to multiple interactions and event
pile-up, calorimeter noise effects, and the energy response of the
calorimeter.  The calorimeter response was measured from the $p_T$
imbalance in photon + jet events.  The relative uncertainty on
the jet energy calibration is $\approx7$\% for jets with
$20<p_T<250\,\text{GeV}$.  The sensitivity of the measurement to this
calibration was reduced by normalizing the $\Dphi$ distribution to the
integrated dijet cross section.  
Nevertheless, this provides the
largest contribution to the systematic uncertainty ($<7\%$ for
$\Dphi>4\pi/5$ but up to $23\%$ for $\Dphi<2\pi/3$).

\begin{figure}
\includegraphics[scale=0.96]{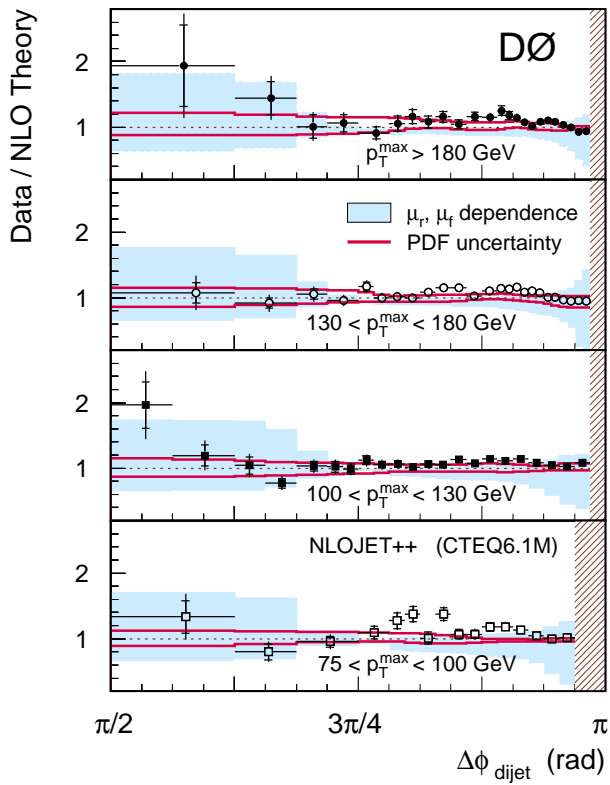}
\caption{\label{fig:data2} 
Ratios of data to the NLO pQCD calculation
for different regions of $\ptmax$.  
Theoretical uncertainties due to variation of $\mu_r$ and $\mu_f$
are shown as the shaded regions;
the uncertainty due to the PDFs is indicated by the solid lines.
The points at large $\Dphi$ are excluded because the calculation has 
non-physical behavior near the divergence at $\pi$.}
\end{figure}

The correction for migrations between bins due to finite energy and
position resolution was determined from events generated with the
{\sc herwig}~\cite{herwig} and {\sc pythia}~\cite{pythia} programs.
The generated jets were smeared according to detector 
resolutions~\cite{kupco}.
The angular jet resolution was determined from a full 
simulation of the D\O\  detector response.
It was found to be better than 20\,mrad for jets with 
energies above 80\,GeV.
The jet $p_T$ resolution was measured from the $p_T$ imbalance in 
dijet events.
It decreases from 18\% at $p_T = 40\,\text{GeV}$ 
to 9\% for $p_T = 200\,\text{GeV}$.
Finite jet $p_T$ resolution can lead to ambiguities in the
selection of the two leading $p_T$ jets.
This effect is large at small $\Dphi$ where contributions
from higher jet multiplicities dominate.
The generated events were reweighted to describe the observed 
$\Dphi$ distribution.
This provided a good description of the observed $p_T$ spectra of the 
four leading $p_T$ jets.
The correction for migrations is typically less than $8\%$ for 
$\Dphi>2\pi/3$ and $\approx 40\%$ for $\Dphi \approx \pi/2$
with  a model dependence of less than $2\%$.
Only for $\ptmax < 130\,\text{GeV}$ and at $\Dphi \approx \pi/2$, 
is the model dependence as large as $\approx 14\%$.

%
The corrected data are presented in Fig.~\ref{fig:data1} as a function
of $\Dphi$ in four ranges of $\ptmax$. 
The inner error bars represent the statistical uncertainties and the 
outer error bars correspond to the quadratic sum of the statistical and 
systematic uncertainties.  
The spectra are strongly peaked at $\Dphi\approx\pi$; 
the peaks are narrower at larger values of $\ptmax$.
Overlaid on the data points in Fig.~\ref{fig:data1} are the results of pQCD
calculations obtained using the parton-level event generator {\sc
nlojet++}~\cite{nlojet} and  CTEQ6.1M~\cite{cteq6} parton distribution
functions (PDFs) with $\alpha_s(M_Z)=0.118$.
The observable was calculated from the ratio of the predictions
for $2\rightarrow 3$ processes  ($ d \sigma_{\text{dijet}}/ d \Dphi$)
and  $2\rightarrow 2$ processes ($\sigma_{\text{dijet}}$),
both at leading order (LO) or NLO, 
\[
\frac{1}{\sigma_{\text{dijet}}} \Biggr|_{\text{(N)LO}} 
\;
\frac{ d \sigma_{\text{dijet}}}{ d \Dphi}
\Biggr|_{\text{(N)LO}}   \; .
\]
The renormalization and factorization scales are chosen to be
$\mu_r=\mu_f=0.5\,\ptmax$.
The ratio is insensitive to hadronization corrections and the
underlying event~\cite{mwhcp}.

NLO pQCD provides a good description of the data.  
As shown in Fig.~\ref{fig:data2} data and NLO agree within 5--10\%.
The theoretical uncertainty due to the PDFs~\cite{cteq6} is estimated to 
be below 20\%.
Also shown is the effect of renormalization and factorization scale variation
($0.25\,\ptmax < \mu_{r,f} < \ptmax$).  
The large scale dependence for $\Dphi<2\pi/3$ occurs because the 
NLO calculation only receives contributions from tree-level 
four-parton final states in this regime.
Results from pQCD at large $\Dphi$ in 
Figs.~\ref{fig:data1} and~\ref{fig:data2} were 
excluded because fixed-order perturbation theory fails to describe the
data in the region $\Delta\phi_{dijet}\approx\pi$ where soft processes
dominate.

\begin{figure}
\includegraphics[scale=0.96]{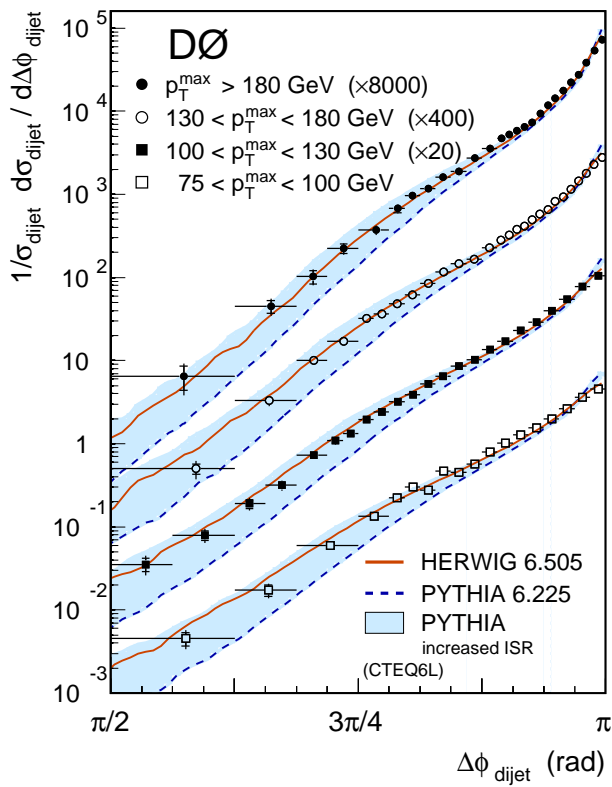}
\caption{\label{fig:data3} 
The $\Dphi$ distributions in different $\ptmax$ ranges.  
Results from {\sc herwig} and {\sc pythia} are overlaid on the data.
Data and predictions with $\ptmax > 100\,\text{GeV}$ are scaled by 
successive factors of 20 for purposes of presentation.}
\end{figure}

Monte Carlo event generators, such as {\sc herwig} and {\sc pythia},
use $2 \rightarrow 2$  LO pQCD matrix elements with phenomenological 
parton-shower models to simulate higher order QCD effects.  
Results from {\sc herwig} (version 6.505) and {\sc pythia} (version 6.225),
both using default parameters and the CTEQ6L~\cite{cteq6} PDFs, are compared
to the data in Fig.~\ref{fig:data3}.
{\sc herwig} 
describes the data well over the 
entire $\Dphi$ range including $\Dphi\approx\pi$.
{\sc pythia} with default parameters describes the data poorly---the 
distribution is too narrowly peaked at $\Dphi\approx\pi$ and lies
significantly below the data over most of the $\Dphi$ range.
The maximum $p_T$ in the initial-state parton shower is directly
related to the maximum virtuality that can be adjusted in {\sc pythia}.
The shaded bands in Fig.~\ref{fig:data3} indicate the range of variation
when the maximum allowed virtuality is smoothly increased from the
current default by a factor of four~\cite{parp67}.
These variations result in significant changes in the low $\Dphi$ region
clearly demonstrating the sensitivity of this measurement.
Consequently, global efforts to tune Monte Carlo event generators 
should benefit from including our data.

To summarize, 
we have measured the dijet azimuthal decorrelation in different ranges
of leading jet $p_T$ and observe an increased decorrelation towards
smaller $p_T$.  NLO pQCD describes the data except for very large
$\Dphi$ where the calculation is not predictive.

We wish to thank W. Giele, Z. Nagy, M.~H. Seymour,
and T. Sj\"ostrand for many helpful discussions.
\input acknowledgement_paragraph_r2.tex
\end{document}

%% file: list_of_authors_r2.tex
%
\author{                                                                      
V.M.~Abazov,$^{33}$                                                           
B.~Abbott,$^{70}$                                                             
M.~Abolins,$^{61}$                                                            
B.S.~Acharya,$^{27}$                                                          
D.L.~Adams,$^{68}$                                                            
M.~Adams,$^{48}$                                                              
T.~Adams,$^{46}$                                                              
M.~Agelou,$^{17}$                                                             
J.-L.~Agram,$^{18}$                                                           
S.N.~Ahmed,$^{32}$
% ...
\\                                                                            
\vskip 0.30cm                                                                 
\centerline{(D\O\ Collaboration)}                                             
\vskip 0.30cm                                                                 
}                                                                             
\address{                                                                     
\centerline{$^{1}$Universidad de Buenos Aires, Buenos Aires, Argentina}       
\centerline{$^{2}$LAFEX, Centro Brasileiro de Pesquisas F{\'\i}sicas,         
                  Rio de Janeiro, Brazil}                                     
\centerline{$^{3}$Universidade do Estado do Rio de Janeiro,                   
                  Rio de Janeiro, Brazil}                                     
\centerline{$^{4}$Instituto de F\'{\i}sica Te\'orica, Universidade            
                  Estadual Paulista, S\~ao Paulo, Brazil}                     
\centerline{$^{5}$University of Alberta and Simon Fraser University,          
                  Canada}                                                     
\centerline{$^{6}$Institute of High Energy Physics, Beijing,                  
                  People's Republic of China}                                 
\centerline{$^{7}$Universidad de los Andes, Bogot\'{a}, Colombia}             
\centerline{$^{8}$Charles University, Center for Particle Physics,            
                  Prague, Czech Republic}                                     
\centerline{$^{9}$Czech Technical University, Prague, Czech Republic}         
\centerline{$^{10}$Institute of Physics, Academy of Sciences, Center          
                  for Particle Physics, Prague, Czech Republic}               
\centerline{$^{11}$Universidad San Francisco de Quito, Quito, Ecuador}        
\centerline{$^{12}$Laboratoire de Physique Corpusculaire, IN2P3-CNRS,         
                 Universit\'e Blaise Pascal, Clermont-Ferrand, France}        
\centerline{$^{13}$Laboratoire de Physique Subatomique et de Cosmologie,      
                  IN2P3-CNRS, Universite de Grenoble 1, Grenoble, France}     
\centerline{$^{14}$CPPM, IN2P3-CNRS, Universit\'e de la M\'editerran\'ee,     
                  Marseille, France}                                          
\centerline{$^{15}$Laboratoire de l'Acc\'el\'erateur Lin\'eaire,              
                  IN2P3-CNRS, Orsay, France}                                  
\centerline{$^{16}$LPNHE, Universit\'es Paris VI and VII, IN2P3-CNRS,         
                  Paris, France}                                              
\centerline{$^{17}$DAPNIA/Service de Physique des Particules, CEA, Saclay,    
                  France}                                                     
\centerline{$^{18}$IReS, IN2P3-CNRS, Univ. Louis Pasteur Strasbourg,          
                   and Univ. de Haute Alsace, France}                         
\centerline{$^{19}$Institut de Physique Nucl\'eaire de Lyon, IN2P3-CNRS,      
                   Universit\'e Claude Bernard, Villeurbanne, France}         
\centerline{$^{20}$RWTH Aachen, III. Physikalisches Institut A,               
                   Aachen, Germany}                                           
\centerline{$^{21}$Universit{\"a}t Bonn, Physikalisches Institut,             
                  Bonn, Germany}                                              
\centerline{$^{22}$Universit{\"a}t Freiburg, Physikalisches Institut,         
                  Freiburg, Germany}                                          
\centerline{$^{23}$Universit{\"a}t Mainz, Institut f{\"u}r Physik,            
                  Mainz, Germany}                                             
\centerline{$^{24}$Ludwig-Maximilians-Universit{\"a}t M{\"u}nchen,            
                   M{\"u}nchen, Germany}                                      
\centerline{$^{25}$Fachbereich Physik, University of Wuppertal,               
                   Wuppertal, Germany}                                        
\centerline{$^{26}$Panjab University, Chandigarh, India}                      
\centerline{$^{27}$Tata Institute of Fundamental Research, Mumbai, India}     
\centerline{$^{28}$University College Dublin, Dublin, Ireland}                
\centerline{$^{29}$Korea Detector Laboratory, Korea University,               
                   Seoul, Korea}                                              
\centerline{$^{30}$CINVESTAV, Mexico City, Mexico}                            
\centerline{$^{31}$FOM-Institute NIKHEF and University of                     
                  Amsterdam/NIKHEF, Amsterdam, The Netherlands}               
\centerline{$^{32}$University of Nijmegen/NIKHEF, Nijmegen, The               
                  Netherlands}                                                
\centerline{$^{33}$Joint Institute for Nuclear Research, Dubna, Russia}       
\centerline{$^{34}$Institute for Theoretical and Experimental Physics,        
                  Moscow, Russia}                                             
\centerline{$^{35}$Moscow State University, Moscow, Russia}                   
\centerline{$^{36}$Institute for High Energy Physics, Protvino, Russia}       
\centerline{$^{37}$Petersburg Nuclear Physics Institute,                      
                   St. Petersburg, Russia}                                    
\centerline{$^{38}$Lund University, Royal Institute of Technology,            
                   Stockholm University, and Uppsala University, Sweden}      
\centerline{$^{39}$Lancaster University, Lancaster, United Kingdom}           
\centerline{$^{40}$Imperial College, London, United Kingdom}                  
\centerline{$^{41}$University of Manchester, Manchester, United Kingdom}      
\centerline{$^{42}$University of Arizona, Tucson, Arizona 85721}              
\centerline{$^{43}$Lawrence Berkeley National Laboratory and University of    
                  California, Berkeley, California 94720}                     
\centerline{$^{44}$California State University, Fresno, California 93740}     
\centerline{$^{45}$University of California, Riverside, California 92521}     
\centerline{$^{46}$Florida State University, Tallahassee, Florida 32306}      
\centerline{$^{47}$Fermi National Accelerator Laboratory, Batavia,            
                   Illinois 60510}                                            
\centerline{$^{48}$University of Illinois at Chicago, Chicago,                
                   Illinois 60607}                                            
\centerline{$^{49}$Northern Illinois University, DeKalb, Illinois 60115}      
\centerline{$^{50}$Northwestern University, Evanston, Illinois 60208}         
\centerline{$^{51}$Indiana University, Bloomington, Indiana 47405}            
\centerline{$^{52}$University of Notre Dame, Notre Dame, Indiana 46556}       
\centerline{$^{53}$Iowa State University, Ames, Iowa 50011}                   
\centerline{$^{54}$University of Kansas, Lawrence, Kansas 66045}              
\centerline{$^{55}$Kansas State University, Manhattan, Kansas 66506}          
\centerline{$^{56}$Louisiana Tech University, Ruston, Louisiana 71272}        
\centerline{$^{57}$University of Maryland, College Park, Maryland 20742}      
\centerline{$^{58}$Boston University, Boston, Massachusetts 02215}            
\centerline{$^{59}$Northeastern University, Boston, Massachusetts 02115}      
\centerline{$^{60}$University of Michigan, Ann Arbor, Michigan 48109}         
\centerline{$^{61}$Michigan State University, East Lansing, Michigan 48824}   
\centerline{$^{62}$University of Mississippi, University, Mississippi 38677}  
\centerline{$^{63}$University of Nebraska, Lincoln, Nebraska 68588}           
\centerline{$^{64}$Princeton University, Princeton, New Jersey 08544}         
\centerline{$^{65}$Columbia University, New York, New York 10027}             
\centerline{$^{66}$University of Rochester, Rochester, New York 14627}        
\centerline{$^{67}$State University of New York, Stony Brook,                 
                   New York 11794}                                            
\centerline{$^{68}$Brookhaven National Laboratory, Upton, New York 11973}     
\centerline{$^{69}$Langston University, Langston, Oklahoma 73050}             
\centerline{$^{70}$University of Oklahoma, Norman, Oklahoma 73019}            
\centerline{$^{71}$Brown University, Providence, Rhode Island 02912}          
\centerline{$^{72}$University of Texas, Arlington, Texas 76019}               
\centerline{$^{73}$Rice University, Houston, Texas 77005}                     
\centerline{$^{74}$University of Virginia, Charlottesville, Virginia 22901}   
\centerline{$^{75}$University of Washington, Seattle, Washington 98195}       
}                                                                             

%% file: acknowledgement_paragraph_r2.tex
%
We thank the staffs at Fermilab and collaborating institutions, 
and acknowledge support from the 
Department of Energy and National Science Foundation (USA),  
Commissariat  \` a l'Energie Atomique and 
CNRS/Institut National de Physique Nucl\'eaire et 
de Physique des Particules (France), 
Ministry of Education and Science, Agency for Atomic 
   Energy and RF President Grants Program (Russia),
CAPES, CNPq, FAPERJ, FAPESP and FUNDUNESP (Brazil),
Departments of Atomic Energy and Science and Technology (India),
Colciencias (Colombia),
CONACyT (Mexico),
KRF (Korea),
CONICET and UBACyT (Argentina),
The Foundation for Fundamental Research on Matter (The Netherlands),
PPARC (United Kingdom),
Ministry of Education (Czech Republic),
Natural Sciences and Engineering Research Council and 
WestGrid Project (Canada),
BMBF and DFG (Germany),
A.P.~Sloan Foundation,
Civilian Research and Development Foundation,
Research Corporation,
Texas Advanced Research Program,
and the Alexander von Humboldt Foundation.

%% file: azimuth.bbl
\begin{thebibliography}{99}
%
\bibitem[*]{lehner}
Visitor from University of Zurich, Zurich, Switzerland.
\bibitem[\dag]{przybycien}
Visitor from Institute of Nuclear Physics, Krakow, Poland.
%
\vskip 0.25cm


\bibitem{multiparton}  M. Dobbs {\it et al.}, in
{\sl Report of the Working Group on Quantum Chromodynamics and the 
      Standard Model,
      3rd Les Houches Workshop on Physics at TeV Colliders}, Les Houches, 
       France, (2004), \texttt{hep-ph/0403100}.

\bibitem{run2cone} G.~C.~Blazey {\it et al.}, in
     {\sl Proceedings of the Workshop:
     ``QCD and Weak Boson Physics in Run II''},
     edited by U.~Baur, R.~K.~Ellis, and D.~Zeppenfeld, 
     Batavia, Illinois  (2000) p.~47.
     See Section 3.5 for details.

\bibitem{nlo} W.~B.~Kilgore and W.~T.~Giele, in
    {\sl Proceedings of the 35th Rencontres De Moriond}, 
      edited by J. Tran Thanh Van, Les Arcs, France (2000).

\bibitem{nlojet} Z.~Nagy, Phys.~Rev.~Lett.~{\bf 88}, 122003 (2002); \\
                 Z.~Nagy, Phys.~Rev.~D~{\bf 68}, 094002 (2003).

\bibitem{run2det} V.~Abazov {\it et al.} (D\O\ Collaboration),
     in preparation for submission to 
       Nucl.~Instrum.~Methods~Phys.~Res.~A;  \\
     T.~LeCompte and H.~T.~Diehl,
       Ann.~Rev.~Nucl.~Part.~Sci. {\bf 50}, 71 (2000); \\ 
     S. Abachi {\it et al.} (D\O\ Collaboration), 
       Nucl.~Instrum.~Methods~Phys.~Res.~A {\bf 338}, 185 (1994). 

\bibitem{herwig} G.~Marchesini {\it et al.}, 
          Comp.~Phys.~Comm.~{\bf 67}, 465 (1992); 
          G.~Corcella {\it et al.},  JHEP {\bf 0101}, 010 (2001).

\bibitem{pythia} T.~Sj\"ostrand  {\it et al.},
               Comp.~Phys.~Comm.~{\bf 135}, 238 (2001).          

\bibitem{kupco} A.~Kup\v{c}o, Ph.D. thesis, 
           Charles University, Prague, Czech Republic (2003), 
         \texttt{FERMILAB THESIS-2004-08}.

\bibitem{cteq6} J.~Pumplin {\it et al.},  JHEP~{\bf 0207}, 12 (2002); \\
                D.~Stump {\it et al.},   JHEP~{\bf0310}, 046 (2003).

\bibitem{mwhcp} M.~Wobisch, to appear in {\sl Proceedings of 
the 15th Topical Conference on Hadron Collider Physics, HCP2004}.

\bibitem{parp67} The {\sc pythia} parameter \texttt{PARP(67)} was 
    increased from the current default of $1.0$ to $4.0$ which was 
    the default before version 6.138.

\end{thebibliography}
